\documentclass[prx,aps,twocolumn,superscriptaddress]{revtex4}

\usepackage{color}
\usepackage[pdftex]{graphicx}
\usepackage{dcolumn}
\usepackage{bm}
\usepackage{amssymb}

\begin{document}


\title{Dynamic inhomogeneities and phase separation after quantum quenches \\ in strongly correlated systems}

\author{Gia-Wei Chern}
\affiliation{Department of Physics, University of Virginia, Charlottesville, VA 22904, USA}

\date{\today}

\begin{abstract}
We present a Gutzwiller von~Neumann dynamics (GvND)  method for simulating equilibrium and nonequilibrium phenomena in strongly correlated electron systems. Our approach is a real-space formulation of the time-dependent Gutzwiller approximation method. Applying the GvND method to simulate interaction quenches in the Peierls-Hubbard model, we demonstrate the amplification of initial inhomogeneities, which in turn results in the collapse of quench-induced synchronized oscillation. Moreover, we find a dynamical phase transition separating two quasi-stationary regimes with rather distinct spatial distributions of physical quantities after the collapsed oscillation. In particular, in the strong-coupling regime, the system exhibits a dynamic phase separation in the quasi-stationary state. Our results thus underscore the importance of spatial fluctuations in the nonequilibrium dynamics of strongly correlated systems.
\end{abstract}

\maketitle

The recent enormous theoretical interest in nonequilibrium dynamics of correlated materials is partly driven by the impressive progress in experimental techniques for controlling and probing such systems~\cite{polkovnikov11,eisert15,langen15}. In particular, developments of time-resolved measurement techniques in solids and cold atoms now allow one to study dynamical phase transitions far from equilibrium on the microscopic time scale of correlated  systems~\cite{perfetti06,basov11}. Theoretically, the quantum quench setup provides an idealized platform to investigate intrinsic out-of-equilibrium dynamics related to strong electron correlation~\cite{rigol07,cazalilla06,kollath07,kollar08,manmana07,rossini09,moeckel08,eckstein09,tsuji13}. 
One crucial question here is whether the system eventually thermalize and reach a new equilibrium~\cite{rigol08}. After a quench to a large interaction parameter, the system exhibits characteristic collapse-and-revival oscillations which eventually fade out in the long time. But in some cases, the system is trapped in a nonthermal quasi-stationary state for very long times~\cite{manmana07,kollath07,tsuji13}. 

While there has been considerable progress in our understanding of nonequilibrium dynamics in one dimension~\cite{white04,daley04}, investigation of quantum quench in high dimensional systems has only begun recently. Here we are mainly interested in interaction quench in fermionic systems, of which the single-band Hubbard model is a canonical example. The nonequilibrium dynamics of a Fermi sea after a sudden switch-on of the Hubbard repulsion in infinite dimensions has been studied in a pioneering work~\cite{eckstein09} using the time-dependent dynamical mean-field theory (DMFT)~\cite{freericks06,aoki14}.  The results clearly indicate the existence of a dynamical phase transition separating two distinct out-of-equilibrium regimes depending on the Hubbard parameter $U_f$ after quench. 
However, up to date, most theoretical studies of quench dynamics ignores the spatial inhomogeneity and fluctuations, which seem to be ubiquitous in nonequilibrium dynamics of many-body systems. A famous example is the formation of topological defects, described by the Kibble-Zurek mechanism~\cite{kibble76,zurek85}, when a system is driven across a continuous phase transition. The DMFT method, even with its cluster or real-space generalization~\cite{georges96,maier05,kotliar06}, is still very limited in its treatment of complex spatial structures due to its heavy computational cost. It is thus highly desirable to develop an approximate yet efficient approach for large-scale real-space simulations of correlated systems. In this regard, the recently developed time-dependent Gutzwiller approximation (TDGA)~\cite{schiro10,schiro11,sandri13,seibold05} provides such an efficient alternative to the more accurate nonequilibrium DMFT.  More importantly, TDGA was shown to capture many nontrivial effects observed in DMFT~\cite{schiro10}, such as the existence of a dynamical phase transition, when applied to quantum quenches in the Hubbard model.

In this paper, we present a Gutzwiller-von~Neumann dynamics (GvND) formulation for real-space dynamical simulations of correlated systems. In this approach, the time evolution of the Gutzwiller variational parameters is coupled to the von~Neumann equation governing the dynamics of many-electron wavefunction. We apply the GvND method to investigate the interaction quench in a triangular-lattice Hubbard model with electron-phonon coupling and uncover several dramatic effects caused by the dynamical lattice degrees of freedom. First, the phase-locked oscillation of, e.g. double occupation, caused by the sudden turn-on of interaction is disrupted by the development of inhomogeneous lattice deformations. Moreover, we find two regimes of quasi-steady states with rather distinct spatial distributions of physical quantities after the collapsed oscillation. In particular, for large $U_f$, the system exhibits a spontaneous phase separation after the quantum quench. 




We consider the Hubbard model on a deformable triangular lattice described by the Hamiltonian
\begin{eqnarray}
	\label{eq:H}
	\mathcal{H} &=& \sum_{\langle ij \rangle, \,\alpha } t(\mathbf u_i - \mathbf u_j) c^\dagger_{i,\alpha} c^{\;}_{j, \alpha}  + U \sum_i n_{i,\uparrow} n_{i,\downarrow} \nonumber \\
		& & + \frac{K}{2} \sum_{\langle ij \rangle} \left[\hat{\mathbf e}_{ij} \cdot (\mathbf u_j - \mathbf u_i) \right]^2 + \sum_i \frac{|\mathbf p_i|^2}{2m}.
\end{eqnarray}
Here $c^\dagger_{i, \alpha}$ is the creation operator of electron with spin $\alpha = \uparrow, \downarrow$ at site-$i$, $n_{i, \alpha} = c^\dagger_{i,\alpha} c^{\;}_{i, \alpha}$ is the electron number operator, $U$ is the Hubbard repulsion parameter, $K$ is a elastic constant, $\mathbf p_i$ is the momentum operator, and $m$ is the mass of the atom. $\mathbf u_i$ denotes the displacement vector of the $i$-th site, i.e. $\mathbf r_i = \mathbf r^{(0)}_i + \mathbf u_i$, and $\hat{\mathbf e}_{ij}$ is a unit vector pointing from site-$i$ to $j$. We consider the following dependence of hopping integral on the displacements 
\begin{eqnarray}
	\label{eq:t_ij}
	t_{ij} = t(\mathbf u_i - \mathbf u_j) = t^{(0)}_{ij} \left[1 + g \, \hat{\mathbf e}_{ij} \cdot (\mathbf u_j - \mathbf u_i) \right],
\end{eqnarray}
where $t^{(0)}_{ij}$ is the bare hopping constant, and $g$ is the electron-phonon coupling. The Hamiltonian~(\ref{eq:H}) with $t_{ij}$ given by Eq.~(\ref{eq:t_ij}) is also called the Peierls-Hubbard (PH) model~\cite{mazumdar83,hirsch83}. The 1D version of the PH model is the famous Su-Schrieffer-Heeger model~\cite{su80} with Hubbard interaction. Earlier interest on the 2D square-lattice PH model is motivated by the interest in high-$T_c$ superconductors~\cite{tang88,fehske92,yuan02}. The PH model has served as the basic platform for investigating the interplay of Peierls instability and electron correlation. Since our main interest here is correlation related nonequilibrium phenomena {\em without} spontaneous symmetry-breaking, we focus on half-filled triangular-lattice PH model in which the lack of Fermi surface nesting prevents the system from weak-coupling instability either in the elastic or electronic subsystems. The phonons here are mainly used as a way to introduce dynamic inhomogeneities to the system.

We first briefly review the TDGA method~\cite{schiro10,schiro11}, which can be derived from the Dirac-Frenkel variational principle~\cite{dirac30,frenkel34}. The Gutzwiller wavefunction is expressed as $|\Psi_G(t) \rangle = \hat{\mathcal{P}}_G(t) |\Psi_S(t) \rangle$, where $|\Psi_S(t) \rangle$ is a time-dependent Slater determinant constructed from a renormalized Hamiltonian. $\hat{\mathcal{P}}_G = \hat{\mathcal{P}}_G(\{\hat\Phi_i(t)\})$ is a time-varying Gutzwiller operator, which can be parameterized by a set of variational matrices $\hat\Phi_i$ in the basis of local Hilbert space. This GA formulation is intimately related to the slave-boson approach~\cite{kotliar86,li89}. In a sense, the $\hat\Phi_i$ matrix elements can be viewed as amplitudes of the slave boson coherent state~\cite{lanata17,behrmann16}. 
Importantly, expectation value of local operator is now expressed as a trace: $\langle \Psi_G | \hat{\mathcal{O}} | \Psi_G \rangle = {\rm Tr}(\hat{\mathcal{O}} \,\hat{\Phi})$~\cite{lanata17}. 
The time evolution of  $\Psi_S$ is governed by the Schr\"odinger equation $i \partial_t |\Psi_S \rangle = \mathcal{H}_{\rm GA}[\hat\Phi] |\Psi_S\rangle$, while the variational matrix obeys the equation of motion: $i \partial_t \hat{\Phi}_i = \hat{\mathcal{U}} \hat\Phi_i + \partial \langle \Psi_S| \mathcal{H}_{\rm GA} |\Psi_S \rangle / \partial \hat{\Phi}_i^\dagger$, where $\hat\mathcal{U}$ is the on-site Coulomb interaction expressed in the local basis of $\hat\Phi$, and the electron binding energy is $\langle \Psi_S| \mathcal{H}_{\rm GA} |\Psi_S \rangle = \sum_{ij} t_{ij} \mathcal{R}_{i, \gamma\alpha} \mathcal{R}^*_{j, \gamma\beta} \rho_{j\beta, i\alpha}$. Here we have introduced the reduced electron density matrix $\rho_{j\beta, i\alpha} \equiv \langle c^\dagger_{i, \alpha} c^{\;}_{j, \beta} \rangle$, and $\mathcal{R}_{i, \alpha\beta}$ is the Gutzwiller renormalization factor~\cite{lanata17} 
\begin{eqnarray}
	\mathcal{R}_{i, \alpha\beta}[\hat{\Phi}_i] = {\rm Tr}\left[\hat{\Phi}_i^\dagger \, \hat{c}^\dagger_{i,\alpha} \hat{\Phi}^{\;}_i \hat{c}^{\;}_{i, \beta} \right] \Big/ \sqrt{n_{i\beta} (1-n_{i\beta})}.
\end{eqnarray}
In our real-space formulation, the dynamical equation for the slave bosons reads
\begin{eqnarray}
	\label{eq:dPhidt}
	\frac{d\hat{\Phi}_i}{dt} &=& 
	i  \biggl( \frac{\partial \mathcal{R}_{i, \gamma\alpha}}{\partial \hat\Phi_i^\dagger} \sum_j t_{ij} \mathcal{R}_{j, \gamma\beta}^* \,\rho^{\;}_{j\beta, i\alpha} + \mbox{h.c.} \biggr) \nonumber \\
	& & \,\, -i U \hat{\mathcal{D}} \,\hat{\Phi}_i  -i \mu_{i, \alpha} \,\hat{\mathcal{N}}_{\alpha} \, \hat{\Phi}_i,
\end{eqnarray}
where $\hat\mathcal{N}$ and  $\hat\mathcal{D}$ are the number and double-occupation operators, respectively, in the local basis, and $\mu_i$ are effective chemical potentials determined from the energy conservation condition. The above Eq.~(\ref{eq:dPhidt}) indicates that the evolution of $\hat\Phi_i$ depends on the electron density matrix, whose dynamics is governed by the von~Neumann equation $d\rho/dt = i [\rho, \mathcal{H}_{\rm GA}]$, or explicitly:
\begin{eqnarray}
	\label{eq:dDdt}
	& & \frac{d\rho^{\;}_{i\alpha, j\beta}}{dt} =  i (\epsilon_j - \epsilon_i) \, \rho^{\;}_{i\alpha, j\beta} \\
	& & \quad +  i \sum_k \left(t_{ik} \mathcal{R}^{\;}_{i, \delta \alpha} \mathcal{R}_{k, \delta \gamma}^* \, \rho^{\;}_{k \gamma, j\beta} 
	- \mathcal{R}_{j, \delta\beta}^* \mathcal{R}^{\;}_{k, \delta\gamma} \rho^{\;}_{i \alpha, k \gamma} \, t_{kj} \right) \nonumber 
\end{eqnarray}
Here we have included an diagonal on-site potential $\epsilon_i$ to the Hubbard Hamiltonian. Eqs.~(\ref{eq:dPhidt}) and~(\ref{eq:dDdt}) comprise a complete set of differential equations for the dynamics of correlated electrons in the real-space TDGA framework. We note in passing that the above formulation can be directly generalized to multi-orbital Hubbard-like models by treating $\alpha$, $\beta$, $\cdots$ as the spin-orbital indices. 

In the presence of dynamical lattice degrees of freedom, these two equations are further coupled to the equation of motion of the displacement field:
\begin{eqnarray}
	\label{eq:newton}
	& & m \frac{d^2 \mathbf u_i}{dt^2} = -K\,\sum_j \hat{\mathbf e}_{ij}  \cdot  (\mathbf u_i - \mathbf u_j) \\
	& & \quad\quad + \sum_j \hat{\mathbf e}_{ij} \left[ g \, t^{(0)}_{ij} \mathcal{R}_{i, \gamma\alpha} \mathcal{R}^*_{j, \gamma\beta} \rho_{j\beta, i\alpha} + \mbox{h.c.} \right]. \nonumber 
\end{eqnarray}
This equation is basically the Newton equation for the atoms, and the above formulation can be viewed as a {\em non-adiabatic} generalization of the Gutzwiller molecular dynamics (MD) method discussed in Ref.~\cite{chern17}. While the motion of atoms is confined in the vicinity of the lattice points here, Ref.~\cite{chern17} considers full MD dynamics for strongly correlated systems in the liquid phase in the {\em adiabatic limit}, which assumes that electrons quickly relax to the equilibrium state of the instantaneous renormalized Hamiltonian. Consequently, both $\hat\Phi_i$ and $\rho_{i\alpha, j\beta}$ are determined self-consistently from the ionic configuration at each time step~\cite{chern17}. In our GvND formulation here, both the slave bosons and the electrons follow their own dynamics in the more general non-adiabatic and non-equilibrium situation.


We next apply the above GvND dynamics method to simulate interaction quench in the triangular-lattice PH model. For single-band Hubbard system, the most general form of $\hat\Phi_i$ is a diagonal matrix with diagonal elements $\Phi_0$ for empty site, $\Phi_{\uparrow}$ and $\Phi_{\downarrow}$, for singly occupied state with a spin-up or down electron, and finally $\Phi_{\uparrow\downarrow}$ for a doubly occupied site. In this basis, both the double-occupation and number operators are also diagonal: $\hat\mathcal{D} = {\rm diag}[0, 0, 0, 1]$, $\hat\mathcal{N}_\uparrow = {\rm diag}[0, 1, 0, 1]$, and $\hat{\mathcal{N}}_\downarrow = {\rm diag}[0,0,1,1]$. Here we focus on non-magnetic electronic states, and assume $n_{i, \uparrow} = n_{i, \downarrow}$ and $\Phi_{i, \uparrow} = \Phi_{i, \downarrow}$, $\rho_{i\alpha, j\beta} = \delta_{\alpha\beta} \rho_{ij}$ and $\mathcal{R}_{i, \alpha\beta} = \delta_{\alpha\beta} \mathcal{R}_i$. In this special case, we have $\mu_{i,\uparrow} = \mu_{i,\downarrow} = {\rm Re}(\sum_j \mathcal{R}_i  \mathcal{R}_j t_{ij} \rho_{ij} ) (n_i - \frac{1}{2})/[n_i (1-n_i)]$ for the chemical potentials.

\begin{figure}[t]
\includegraphics[width=0.95\columnwidth]{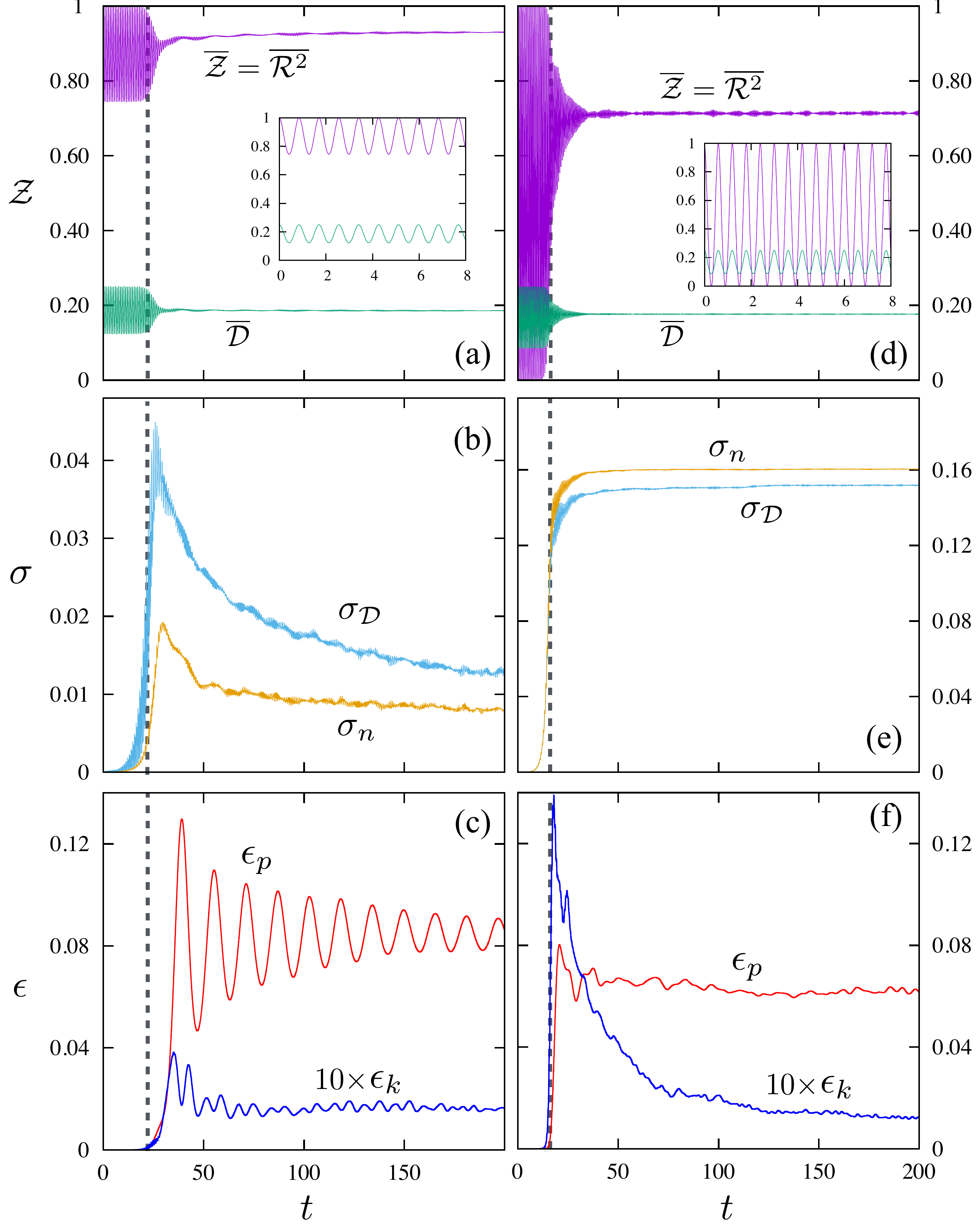}
\caption{(Color online)  
\label{fig:cmp1} GvND dynamics simulation of quantum quench with a $U_f = 0.4\,W$ (left) and $1.33 \, W$ (right). The various panels show the time dependence of averaged quasiparticle weight $\overline{\mathcal{Z}}$ and double-occupation $\overline{\mathcal{D}}$ [(a) and (d)], the standard deviation of double occupation and on-site density [(b) and (e)], and the elastic potential $\epsilon_p$ and kinetic $\epsilon_k$ energy densities [(c) and (f)]. 
}
\end{figure}

In our interaction quench simulations, the system is initially in the $T = 0$ ground state of a hal-filled PH model with $U = 0$. At $t > 0$, the Hubbard repulsion is suddenly turned on to $U_f > 0$. Fig.~\ref{fig:cmp1} shows the time dependence of the various physical quantities from simulations with $U_f= 0.4\, W$ and $1.33\, W$, where $W = 9\,t_{\rm nn}$ is the bandwidth. The parameters used in these simulations are $g = 0.64$, $K = 0.5$, and mass $m = 2$; energy and inverse time are measured in units of the nearest-neighbor hopping $t_{\rm nn}$. After the Coulomb interaction is switched on, the spatial averaged double-occupation $\overline \mathcal{D}$ and the quasiparticle weight $\overline\mathcal{Z} = \overline{\mathcal{R}^2}$ exhibits the characteristic oscillation in short time scales, as shown in the insets of Fig.~\ref{fig:cmp1}(a) and~(d). Such oscillations have been reported previously in DMFT~\cite{eckstein09} as well as TDGA~\cite{schiro10} simulations of the Hubbard model.

Interestingly, we find that this oscillation only lasts up to a time $t^*$, beyond which both the quasiparticle weight and double-occupation  approach  quasi-stationary value $\mathcal{Z}^*$ and $\mathcal{D}^*$, respectively, after a transient period. By investigating the behavior of individual sites, we find that both quantities continue to oscillate but with a {\em site-dependent} amplitude and frequency. This phenomenon can thus be viewed as the collapse of synchronized oscillations among individual sites. To understand what causes this collapse, we note that the collapse at $t^*$ coincides with the rapid rise of the standard deviations, $\sigma_{\mathcal{D}}$ and $\sigma_n$, of double-occupation and on-site density, indicating the development of a inhomogeneous configuration.

\begin{figure}[t]
\includegraphics[width=0.99\columnwidth]{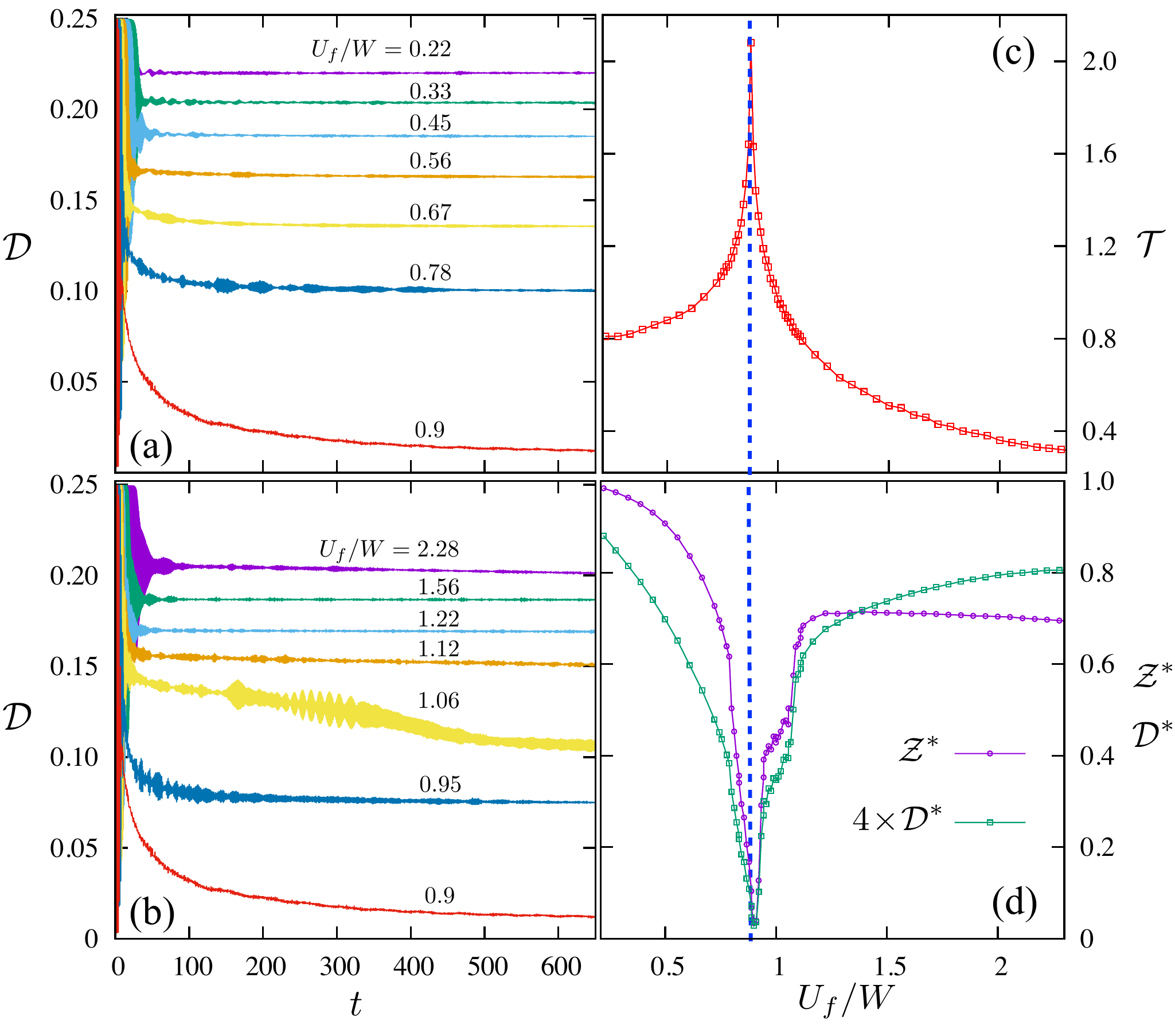}
\caption{(Color online)  
\label{fig:U-dep} (a) and (b) show the time dependence of averaged double occupation for varying $U_f$. (c) the period $\mathcal{T}$ of coherent oscillation (before collapse) as a function of $U_f$. (d) the quasi-stationary $\mathcal{Z}^*$ and $\mathcal{D}^*$ vs $U_f$
}
\end{figure}

\begin{figure}[t]
\includegraphics[width=0.99\columnwidth]{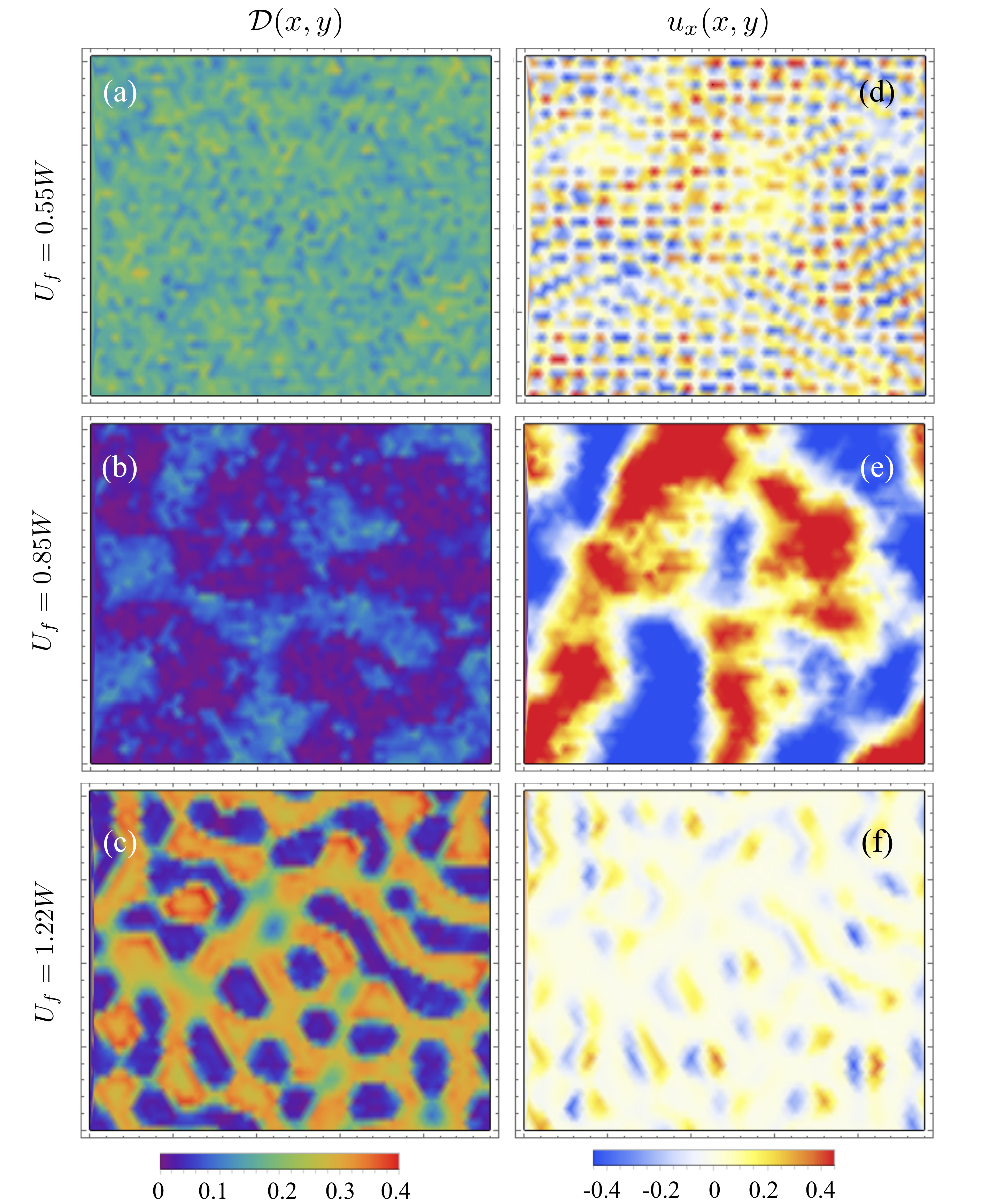}
\caption{(Color online)  
\label{fig:snapshots} Snapshots of the double-occupation $\mathcal{D}(x,y)$ and displacement field $u_x(x, y)$ at a time $t \approx 50 t^*$ for varying $U_f$. The lattice size is $48\times 48$.
}
\end{figure}

It is worth noting that the coherent oscillation after the quench is a unstable quasi-stationary state. We find that the collapse of the oscillation results from the amplification of initial disorder or inhomogeneity, no matter how small it is. In order to have a more controlled simulation, we introduce random displacements of the order of $|\mathbf u_i| \sim 10^{-6}$ in the initial state. The electron contribution to the forces acting on an atom is given by the second term in Eq.~(\ref{eq:newton}). In the coherent oscillation regime, we can approximate the forces as $\mathbf f_i(t) = g\,t_{\rm nn} \overline{\mathcal{R}}^2(t) \sum_j \hat{\mathbf e}_{ij} \,\delta\rho_{ij}$, where $\delta\rho_{ij}$ is deviation from the density matrix of a uniform electronic state. In a perfectly ordered state, the atoms experience no forces coming from the electrons. However, in the presence of disorder, the rapid oscillation of the renormalization factors tends to amplify the effect of inhomogeneity $\delta\rho_{ij}$, hence injecting energy to the phonons. The increased lattice distortion in turn enhances the inhomogeneity in density matrix. This positive feedback eventually leads to the collapse of the coherent oscillation. Indeed, detailed examination shows that after the quench the elastic energy grows exponentially until the time $t^*$. After the collapse of the oscillation, both the potential and kinetic energy relax to quasi-stationary values, as demonstrated in Fig.~\ref{fig:cmp1}(c) and~(f).  

An intriguing phenomenon in the interaction quenches of Hubbard model is the existence of dynamical phase transition that separates the weak and strong coupling regimes. This result was first reported from the DMFT simulation~\cite{eckstein09}, and was later reproduced using the TDGA approach~\cite{schiro10}. Our GvND simulations also find a dynamical transition at $U_f^c \approx 0.85 \, W$. Fig.~\ref{fig:U-dep}(a) and (b) show the time trace of the (spatial) averaged double occupation in the weak and strong coupling regimes, respectively. The period of the coherent oscillation (before the collapse), shown in Fig.~\ref{fig:U-dep}(c) as a function of $U_f$, diverges as $U_f$ approaches the critical value, consistent with previous result. Moreover, both $\mathcal{Z}^*$ and $\mathcal{D}^*$ tend to zero when approaching the transition point, indicating an emergent dynamical Mott insulating state; see Fig.~\ref{fig:U-dep}(d).

\begin{figure}
\includegraphics[width=0.99\columnwidth]{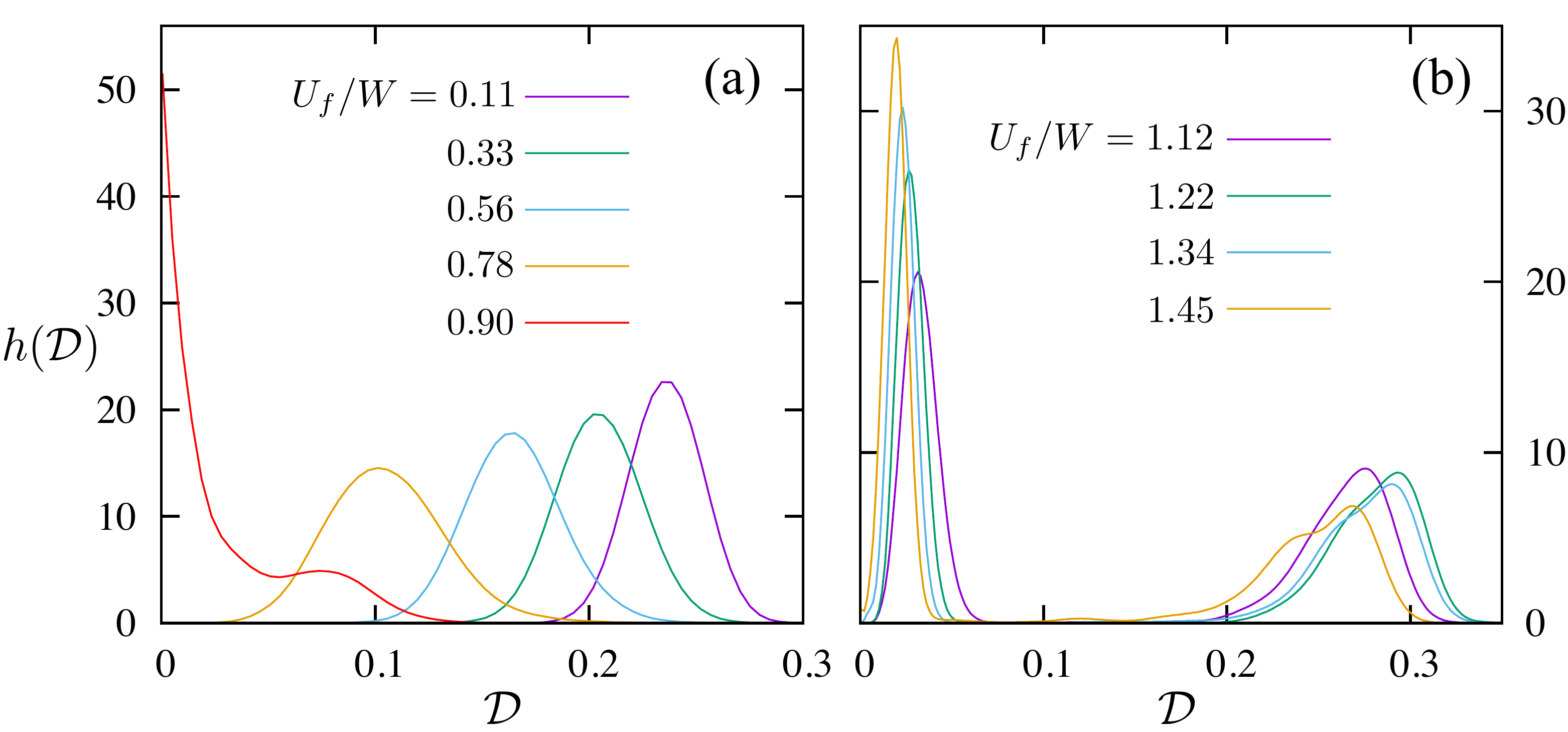}
\caption{(Color online)  
\label{fig:histogram} Probability distribution function of double-occupation in the (a) weak and (b) strong coupling regimes.
}
\end{figure}

The dynamical transition at $U_f^{c}$ not only separates two distinct temporal behaviors as already suggested in previous studies~\cite{eckstein09,schiro10}. Our GvND simulations reveal rather distinct spatial patterns in the quasi-stationary states after the collapse of the coherent oscillation. Fig.~\ref{fig:snapshots} shows the spatial profile of double-occupation and displacement field at a time $t \gg t^*$ for varying $U_f$. In the weak-coupling regime ($U_f < U_f^c$), the distribution of double-occupation is relatively uniform and is dominated by short-range fluctuations. Indeed, its probability distribution $h(\mathcal D)$, shown in Fig.~\ref{fig:histogram}(a), exhibits a single peak with a width that increases as $U_f$ approaches the critical value. Moreover, the displacements clearly shows high-frequency spatial modulation; see Fig.~\ref{fig:snapshots}(d). 
At $U_f \sim U_f^c$, both $\mathcal{D}$ and $u_x$ show fractal-like structures with various length scales, resembling the spatial distribution of order parameters in the vicinity of an equilibrium phase transition. The corresponding $h(\mathcal{D})$  exhibits a pronounced peak at $\mathcal{D} = 0$, consistent with the fact that $\mathcal{D}^* \to 0$ at the dynamical transition point.

On the other hand, a rather distinct distribution is obtained in the strong-coupling regime as demonstrated in Fig.~\ref{fig:snapshots}(c) and~(f). In this regime where $U_f > U_f^c$, domains with large double-occupation and on-site density are interspersed with Mott droplets, or regions with vanishing $\mathcal{D}$ and $\mathcal{Z}$. This picture of phase separation can also be clearly seen from the double-peak structure in the probability distribution of double-occupation in this regime shown in Fig.~\ref{fig:histogram}(b).  Moreover, we find that the Mott droplets are characterized by reduced electron hoppings. This is consistent with the fact that large displacements are found to coincide with these insulating regions, where the hopping amplitude $t_{ij}$ is reduced due to the Peierls coupling; see Eq.~(\ref{eq:t_ij}). This dynamic phase separation in the regime of large $U_f$ is reminiscent of the site-selective Mott transition observed in highly disordered Hubbard model~\cite{byczuk05,aguiar09}. It is found that the insulating phase of the Anderson-Hubbard model with strong on-site disorder is a highly inhomogeneous state with Mott insulating droplets interlaced with regions containing Anderson-localized quasiparticles~\cite{aguiar09}.

To summarize, we have presented a Gutzwiller von~Neumann equation method for simulating real-space nonequilibrium dynamics in strongly correlated electron systems. Applying the GvND method to study the interaction quench in Peierls-Hubbard model, we find that the quench-induced coherent oscillation is intrinsically unstable against any initial disorder.  In fact, this instability against inhomogeneity exists even without the dynamical phonons. A similar collapse of the synchronized oscillation can be induced by a tiny on-site disorder as in the Anderson-Hubbard model~\cite{chern}. In the presence of dynamical lattice degrees of freedom, we further observe dynamically generated phase separation when the system is quenched to a strong-coupling regime. Such amplification of inhomogeneity has already been reported previously in a similar quench dynamics study of the Luttinger liquid~\cite{foster10}. However, in the 1D case, the amplification of the initial inhomogeneity is attributed to fractionalized quasiparticles~\cite{foster10}, which is a special feature of 1D phase. On the other hand, similar dynamical inhomogeneity has also been found in the interaction quenches of high dimensional superfluid~\cite{dzero09}. In our case, the highly inhomogeneous nonequilibrium state certainly results from the nontrivial interplay of correlations and electron-phonon couplings. 

The fact that the elastic part of the Peierls-Hubbard model is a linear system might lead to prolonged post-quench quasi-stationary states since there is no energy exchange between the phonons. Although different phonon modes can talk to each other indirectly through the electrons, we expect inclusion of elastic nonlinearity could introduce additional relaxations and help with the thermalization. Such issues can be addressed with molecular dynamics simulations based on our GvND method, which will be left for future studies. 

\begin{acknowledgements}
The author would like to thank K. Barros, C. Batista, G. Kotliar, and H. Suwa for collaborations on related works and numerous insightful discussions.  This work is partly supported by the Center for Materials Theory as a part of the Computational Materials Science  (CMS) program, funded by the  U.S. Department of Energy, Office of Science, Basic Energy Sciences, Materials Sciences and Engineering Division.
\end{acknowledgements}


\begin{thebibliography}{99}

\bibitem{polkovnikov11} A. Polkovnikov, K. Sengupta, A. Silva, and M. Vengalattore, Nonequilibrium dynamics of closed interacting quantum systems, Rev. Mod. Phys. {\bf 83}, 863 (2011).

\bibitem{eisert15} J. Eisert, M. Friesdorf, and C. Gogolin, Quantum many-body systems out of equilibrium, Nature Phys. {\bf 11}, 124 (2015).

\bibitem{langen15} T. Langen, R. Geiger, and J. Schmiedmayer, Ultracold atoms out of equilibrium, Annu. Rev. Condens. Matter {\bf 6}, 201 (2015).

\bibitem{perfetti06} L. Perfetti, P. A. Loukakos, M. Lisowski, U. Bovensiepen, H. Berger, S. Biermann, P. S. Cornaglia, A. Georges, and M. Wolf, Time Evolution of the Electronic Structure of $1T$-TaS$_2$ through the Insulator-Metal Transition, Phys. Rev. Lett. {\bf 97}, 067402 (2006).

\bibitem{basov11} D. N. Basov, R. D. Averitt, D. van der Marel, M. Dressel, and K. Haule, Electrodynamics of correlated electron materials, Rev. Mod. Phys. {\bf 83}, 471 (2011).

\bibitem{rigol07} M. Rigol, V. Dunjko, V. Yurovsky, and M. Olshanii, Relaxation in a Completely Integrable Many-Body Quantum System: An {\em Ab Initio} Study of the Dynamics of the Highly Excited States of 1D Lattice Hard-Core Bosons, Phys. Rev. Lett. {\bf 98}, 050405 (2007).

\bibitem{cazalilla06} M. A. Cazalilla, Effect of Suddenly Turning on Interactions in the Luttinger Model, Phys. Rev. Lett. {\bf 97}, 156403 (2006).

\bibitem{kollar08} M. Kollar and M. Eckstein, Relaxation of a one-dimensional Mott insulator after an interaction quench, Phys. Rev. A {\bf 78}, 013626 (2008).

\bibitem{rossini09} D. Rossini, A. Silva, G. Mussardo, and G. E. Santoro, Effective Thermal Dynamics Following a Quantum Quench in a Spin Chain, Phys. Rev. Lett. {\bf 102}, 127204 (2009).

\bibitem{eckstein09} M. Eckstein, M. Kollar, and P. Werner, Thermalization after an interaction quench in the Hubbard model, Phys. Rev. Lett. {\bf 103}, 056403 (2009).

\bibitem{moeckel08} M. Moeckel and S. Kehrein, Interaction Quench in the Hubbard Model, Phys. Rev. Lett. {\bf 100}, 175702 (2008).

\bibitem{kollath07} C. Kollath, A. M. L\"auchli, and E. Altman, Quench Dynamics and Nonequilibrium Phase Diagram of the Bose-Hubbard Model, Phys. Rev. Lett. {\bf 98}, 180601 (2007).

\bibitem{manmana07} S. R. Manmana, S. Wessel, R. M. Noack, and A. Muramatsu, Strongly Correlated Fermions after a Quantum Quench, Phys. Rev. Lett. {\bf 98}, 210405 (2007).

\bibitem{tsuji13} N. Tsuji, M. Eckstein, and P. Werner, Nonthermal Antiferromagnetic Order and Nonequilibrium Criticality in the Hubbard Model, Phys. Rev. Lett. {\bf 110}, 136404 (2013).

\bibitem{rigol08} M. Rigol, V. Dunjko, and M. Olshanii, Thermalization and its mechanism for generic isolated quantum systems, Nature {\bf 452}, 854 (2008).


\bibitem{white04} S. R. White and A. E. Feiguin, Real-Time Evolution Using the Density Matrix Renormalization Group, Phys. Rev. Lett. {\bf 93}, 076401 (2004).

\bibitem{daley04} A. J. Daley, C. Kollath, U. Schollw\"ock, and G. Vidal, Time-dependent density-matrix renormalization-group using adaptive effective Hilbert spaces, J. Stat. Mech. P04005 (2004).


\bibitem{freericks06} J. K. Freericks, V. M. Turkowski, and V. Zlati\'c, Nonequilibrium Dynamical Mean-Field Theory, Phys. Rev. Lett. {\bf 97}, 266408 (2006).

\bibitem{aoki14} H. Aoki, N. Tsuji, M. Eckstein, M. Kollar, T. Oka, and P. Werner, Nonequilibrium dynamical mean-field theory and its applications, Rev. Mod. Phys. {\bf 86}, 779 (2014).

\bibitem{kibble76} T. W. B. Kibble, Topology of cosmic domains and strings, J. Phys. A: Math. Gen. {\bf 9}, 1387 (1976).

\bibitem{zurek85} W. H. Zurek, Cosmological experiments in superfluid helium ?, Nature {\bf 317}, 505 (1985).



\bibitem{georges96} A. Georges, G. Kotliar, W. Krauth, and M. J. Rozenberg, Dynamical mean-field theory of strongly correlated fermion systems and the limit of infinite dimensions, Rev. Mod. Phys. {\bf 68}, 13 (1996).

\bibitem{maier05} T. Maier, M. Jarrell, T. Pruschke, and M. H. Hettler, Quantum cluster theories, Rev. Mod. Phys. {\bf 77}, 1027  (2005).

\bibitem{kotliar06} G. Kotliar, S. Y. Savrasov, K. Haule, V. S. Oudovenko, O. Parcollet, and C. A. Marianetti, Electronic structure calculations with dynamical mean-field theory, Rev. Mod. Phys. {\bf 78}, 865 (2006).





\bibitem{schiro10} M. Schir\'o and M. Fabrizio, Time-dependent mean-field theory for quench dynamics in correlated electron systems, Phys. Rev. Lett. {\bf 105}, 076401 (2010).

\bibitem{schiro11} M. Schir\'o and M. Fabrizio, Quantum quenches in the Hubbard model: Time-dependent mean-field theory and the role of quantum fluctuations, Phys. Rev. B {\bf 83}, 165105 (2011).

\bibitem{sandri13} M. Sandri and M. Fabrizio, Nonequilibrium dynamics in the antiferromagnetic Hubbard model, Phys. Rev. B {\bf 88}, 165113 (2013).

\bibitem{seibold05} G. Seibold and J. Lorenzana, Time-Dependent Gutzwiller Approximation for the Hubbard Model, Phys. Rev. Lett. {\bf 86}, 2605 (2005).


\bibitem{mazumdar83} S. Mazumdar and S. N. Dixit, Coulomb Effects on One-Dimensional Peierls Instability: The Peierls-Hubbard Model, Phys. Rev. Lett. {\bf 51}, 292 (1983).

\bibitem{hirsch83} J. E. Hirsch, Effect of Coulomb Interactions on the Peierls Instability, Phys. Rev. Lett. {\bf 51}, 296 (1983).

\bibitem{su80}  W. P. Su, J. R. Schrieffer, and A. J. Heeger, Soliton excitations in polyacetylene, Phys. Rev. B {\bf 22}, 2099 (1980).

\bibitem{tang88} S. Tang and J. E. Hirsch, Peierls instability in the two-dimensional half-filled Hubbard model, Phys. Rev. B {\bf 37}, 9546 (1988).

\bibitem{fehske92} H. Fehske, M. Deeg, and H. B\"uttner, Two-dimensional Peierls-Hubbard model within the slave-boson approach, Phys. Rev. B {\bf 46}, 3713 (1992).

\bibitem{yuan02} Q. Yuan and T. Kopp, Coexistence of the bond-order wave and antiferromagnetism in a two-dimensional hall-filled Peierls-Hubbard model, Phys. Rev. B {\bf 65}, 085102 (2002).


\bibitem{dirac30} P. A. M. Dirac, {\em Note on Exchange Phenomena in the Thomas Atom}, Proc. Cambridge Philos. Soc. {\bf 26}, 376 (1930).

\bibitem{frenkel34} J. Frenkel, {\em Wave Mechanics: Advanced General Theory} (Clarendon, Oxford, 1934).














\bibitem{kotliar86} G. Kotliar and A. E. Ruckenstein, New Functional Integral Approach to Strongly Correlated Fermi Systems: The Gutzwiller Approximation as a Saddle Point, Phys. Rev. Lett. {\bf 57}, 1362 (1986).

\bibitem{li89} T. Li, P. W\"{o}lfle, and P.J. Hirschfeld, Spin-Rotation-Invariant Slave-Boson Approach to the Hubbard Model, Phys. Rev. B {\bf 40}, 6817 (1989).

\bibitem{lanata17} N. Lanat\'{a}, Y. Yao, X. Deng, V. Dobrosavljevi\'{c}, and G. Kotliar, Slave Boson Theory of Orbital Differentiation with Crystal Field Effects: Application to UO$_2$, Phys. Rev. Lett. {\bf 118}, 126401 (2017).

\bibitem{behrmann16} M. Behrmann, A. I. Lichtenstein, M. I. Katsnelson, and F. Lechermann, Versatile approach to spin dynamics in correlated electron systems, Phys. Rev. B {\bf 94}, 165120 (2016).








\bibitem{chern17} G.-W. Chern, K. Barros, C. D. Batista, J. Kress, and G. Kotliar, Mott transition in a metallic liquid -- Gutzwiller molecular dynamics simulations, Phys. Rev. Lett. {\bf 118}, 226401 (2017).







\bibitem{byczuk05} K. Byczuk, W. Hofstetter, and D. Vollhardt, Mott-Hubbard Transition versus Anderson Localization in Correlated Electron Systems with Disorder, Phys. Rev. Lett. {\bf 94}, 056404 (2005).

\bibitem{aguiar09} M. C. O. Aguiar, V. Dobrosavljevi\'c, E. Abrahams, and G. Kotliar, Critical Behavior at the Mott-Anderson Transition: A Typical-Medium Theory Perspective, Phys. Rev. Lett. {\bf 102}, 156402 (2009).


\bibitem{chern} G.-W. Chern, unpublished. 


\bibitem{foster10} M. S. Foster, E. A. Yuzbashyan, and B. L. Altshuler, Quantum Quench in One Dimension: Coherent Inhomogeneity Amplification and ``Supersolitons'', Phys. Rev. Lett. {\bf 105}, 135701 (2010).

\bibitem{dzero09} M. Dzero, E. A. Yuzbashyan, and B. L. Altshuler, Cooper pair turbulence in atomic Fermi gases, Europhys. Lett. {\bf 85}, 20004 (2009).


\end{thebibliography}
\end{document}